\documentclass[prb,twocolumn,showpacs,preprintnumbers,superscriptaddress]{revtex4-1}
\usepackage[usenames,dvipsnames]{color}
\usepackage{amssymb} 
\usepackage{times}
\usepackage{graphicx}
\usepackage{graphicx}
\usepackage{dcolumn}
\usepackage{datetime}
\usepackage{soul}
\usepackage{subfigure}
\usepackage[bookmarks, colorlinks=true, plainpages = false,citecolor = red, urlcolor = black, 
filecolor = blue]{hyperref}
\usepackage{multirow} 
\usepackage[english]{babel}
\usepackage{graphicx}
\usepackage{amsmath}
 
\usepackage[english]{babel}
\usepackage{graphicx}
\usepackage{amsmath}
 
\begin{document}
\title{Strong Spin Filtering by Silicon Nanoparticles with Adsorbed Bismuth Atom Clusters:  Role of Symmetry and Electrostatic Control of the Direction of the Spin Polarization}
 
\author{Alireza Saffarzadeh} 
\affiliation{Department of Physics, Simon Fraser
University, Burnaby, British Columbia, Canada V5A 1S6}
\affiliation{Department of
Physics, Payame Noor University, P.O. Box 19395-3697 Tehran, Iran}
\author{George Kirczenow} 

\affiliation{Department of Physics, Simon Fraser
University, Burnaby, British Columbia, Canada V5A 1S6}

\date{\today}

\begin{abstract}\noindent

We present a theoretical study, based on density functional theory and tight binding modeling, of the electronic structure and spin transport properties of silicon nanoparticles with adsorbed bismuth atoms. We find the bismuth atoms to form clusters separated by quantum tunnel barriers. We predict strong spin filtering by these nanostructures in the high conductance regime when the source and drain leads are connected to the same bismuth cluster and in the low conductance regime when the source and drain leads are connected to different bismuth clusters. We relate the spin filtering to a symmetry obeyed by the spin transmission probability matrix of the system. We also predict that for such systems the direction of the spin polarization in the drain lead can be tuned through large angles and even reversed electrostatically simply by varying the voltage applied to a gate. Realization of these silicon-bismuth nanostructures in the laboratory is feasible. We expect the predicted spin filtering to be experimentally accessible and potentially relevant for device applications.

\end{abstract}

 \maketitle
 
 \section{Introduction}
\label{Intro}

Several experimental and theoretical studies have indicated that {\em planar} monolayer bismuthene on silicon carbide should be a high temperature two-dimensional topological insulator (2DTI).\cite{Hsu2015,Reis2017,Dominguez2018,GLi2018,GK2018,Canonico2019,Azari2019,Hao2019,Stuhler2020}  Building on this work, we have recently studied theoretically the properties of {\em curved} bismuthene-silicon bilayer nanostructures consisting of a monolayer of bismuthene adsorbed on a silicon monolayer.\cite{dome,cowrie} Our density functional theory (DFT) based total energy calculations showed that a hemispherical dome of this kind with a zigzag edge should be stable,\cite{dome} whereas such a dome with an armchair edge should collapse to a cowrie shell-like geometry\cite{cowrie}. Our transport calculations carried out for these structures have predicted that both of these curved systems should be highly effective two-terminal spin filters even in the absence of magnetic fields.\cite{dome,cowrie} By contrast, planar 2DTI spin filters require at least three terminals. However, in the nanostructures considered in Refs. \onlinecite{dome} and \onlinecite{cowrie} the bismuthene monolayer coated the outside of a curved silicon monolayer so that these structures had {\em hollow} interiors, but no plausible methodology for realizing such hollow bismuthene-silicon nanostructures\cite{dome,cowrie} in the laboratory is known at the present time. Here we propose a different class of curved bismuth-on-silicon nanostructures that consist of a bismuth monolayer adsorbed on a solid silicon nanoparticle core. Since silicon nanoparticles ranging in size from 2 to 64 nm have already been synthesized,\cite{Thiessen2019,Huang2021} the experimental realization of such bismuth-on-silicon nanostructures by depositing bismuth atoms on silicon nanoparticles by molecular beam epitaxy should be feasible, and would open the way for pioneering experimental studies of spin transport in curved bismuth monolayer nanostructures. Our calculations  based on DFT and tight binding models predict systems of this type to be nearly perfect two-terminal spin filters in the absence of magnetic fields. We relate this spin filtering to a symmetry obeyed by the spin transmission probability matrix of the system. We also predict that for such systems the direction of the spin polarization of the electric current that exits the spin filter can be tuned through large angles and even reversed simply by varying the voltage applied to an electrostatic gate.

\begin{figure}[b]
\centering
\includegraphics[width=1.0\linewidth]{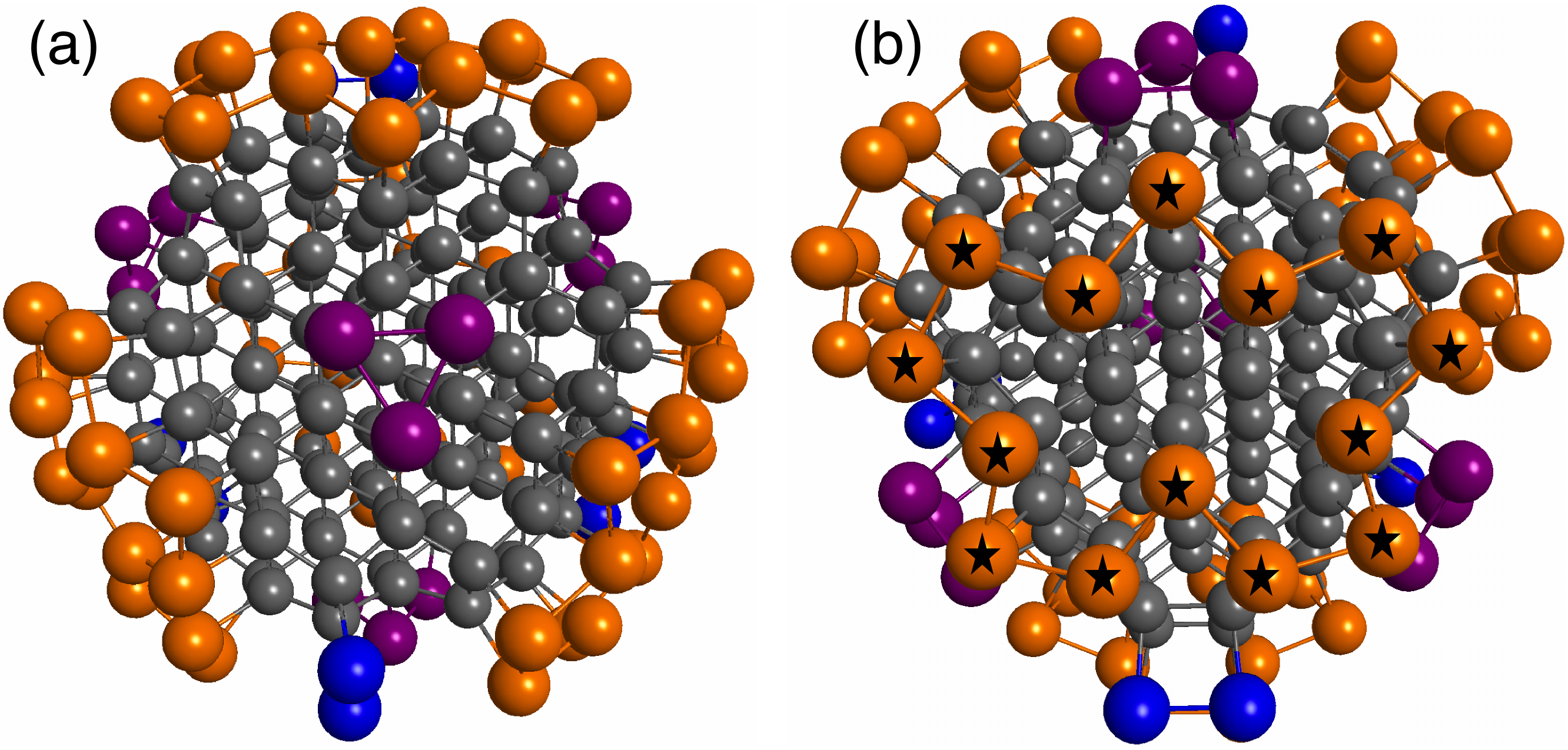}
\caption{(Color online).
Two views of a Bi$_{76}$Si$_{147}$ nanostructure consisting of a silicon core (gray Si atoms) with Bi atoms adsorbed on its surface. The structure has tetrahedral symmetry. The adsorbed Bi atoms form separate clusters of 2 atoms (blue), 3 atoms (purple) and 14 atoms (orange). There are 4 clusters of each type. The bismuth atoms of each 14 atom atom cluster are arranged in a kinked ring geometry; the atoms of one ring are indicated by stars in part (b).  Image prepared using Macmolplt software.\cite{MacMolPlt}}
\label{geometry} 
\end{figure}

\section{Structure}
\label{Structure}

In this paper we consider the Bi$_{76}$Si$_{147}$ nanoparticle shown in Fig.\ref{geometry}. The silicon core of this nanostructure is symmetric under the operations of the tetrahedral point group Td, as is the Bi$_{76}$Si$_{147}$ nanoparticle as a whole. The silicon core has a compact surface geometry with each surface silicon atom having a coordination number of 3 or 4, 4 being the coordination number of the interior silicon atoms. Each bismuth atom bonds in the ``on top" position to a single silicon atom that has coordination number 3; a bismuth atom bonds in this way to each such silicon atom. Thus, for this geometry all of the silicon dangling bonds are saturated by bismuth atoms and consequently, based on conventional considerations of quantum chemistry, it is reasonable to expect the geometry of this Bi$_{76}$Si$_{147}$ nanoparticle to be stable or at least metastable; this expectation was confirmed by our DFT calculations. The structure shown in Fig.\ref{geometry} was relaxed by means of DFT computer simulations. The DFT calculations reported here were carried out  with the GAUSSIAN 16 package using the B3PW91 functional and Lanl2DZ effective core potential and basis sets.\cite{Frisch} The electronic energy and ionic forces of our optimized geometries were converged within 10$^{-5}$ eV and 0.0008 eV/\AA, respectively.
Because of the locations of the silicon atoms with coordination number of 3 on the surface of the silicon core, the adsorbed Bi atoms of this bismuth-on-silicon nanostructure form separate atomic clusters as can be seen for the Bi$_{76}$Si$_{147}$ structure shown in Fig.\ref{geometry}. We note that, unlike for the structure in Fig.\ref{geometry}, in both the bismuthene-silicon bilayer domes and cowrie shell-like nanostructures that were studied previously,\cite{dome,cowrie} the Bi atoms were arranged in a continuous network, not a collection of separate clusters. As will be seen below, the partition of the Bi atoms of the Bi$_{76}$Si$_{147}$ structure considered here into clusters separated by spatial gaps that act as tunnel barriers has a strong effect on both electron transport and spin transport through this nanostructure. In this paper we present results for the Bi$_{76}$Si$_{147}$ structure shown in Fig.\ref{geometry}. However, we have also carried out spin transport calculations for a Bi$_{60}$Si$_{147}$H$_{16}$ nanostructure that differs from our Bi$_{76}$Si$_{147}$ nanostructure in that the bismuth atoms belonging to the small 2 and 3 atom bismuth clusters are replaced with hydrogen atoms and found spin filtering results qualitatively similar to those for our Bi$_{76}$Si$_{147}$ nanostructure.

\section{Tight Binding Model}
\label{M}

Several tight binding models have successfully captured the crucial physics of bismuthene on SiC.\cite{Reis2017,Dominguez2018,GLi2018,GK2018,Canonico2019,Azari2019,Hao2019} These models\cite{Reis2017,Dominguez2018,GLi2018,GK2018,Canonico2019,Azari2019,Hao2019} have employed basis sets consisting {\em only} of the valence orbitals of the bismuth atoms but have been parameterized to take into account the influence of the SiC substrate on the bismuthene. We have extended this approach to construct tight binding models of  curved nanoscale bismuthene-silicon bilayer domes\cite{dome} and cowrie shell-like structures.\cite{cowrie} These are Slater-Koster type models\cite{SK1954} modified to explicitly take into account the curved nature of the bismuthene layer. They include the site dependence of the electron electrostatic potential energy, as well as spin-orbit coupling\cite{PRBrapid,PRB} and Rashba\cite{BR1,BR2} contributions to the Hamiltonian.  The model parameters were adjusted so as to match the bismuthene partial densities of states calculated for these nanostructures within DFT.

\begin{figure}[b]
\centering
\includegraphics[width=0.9\linewidth]{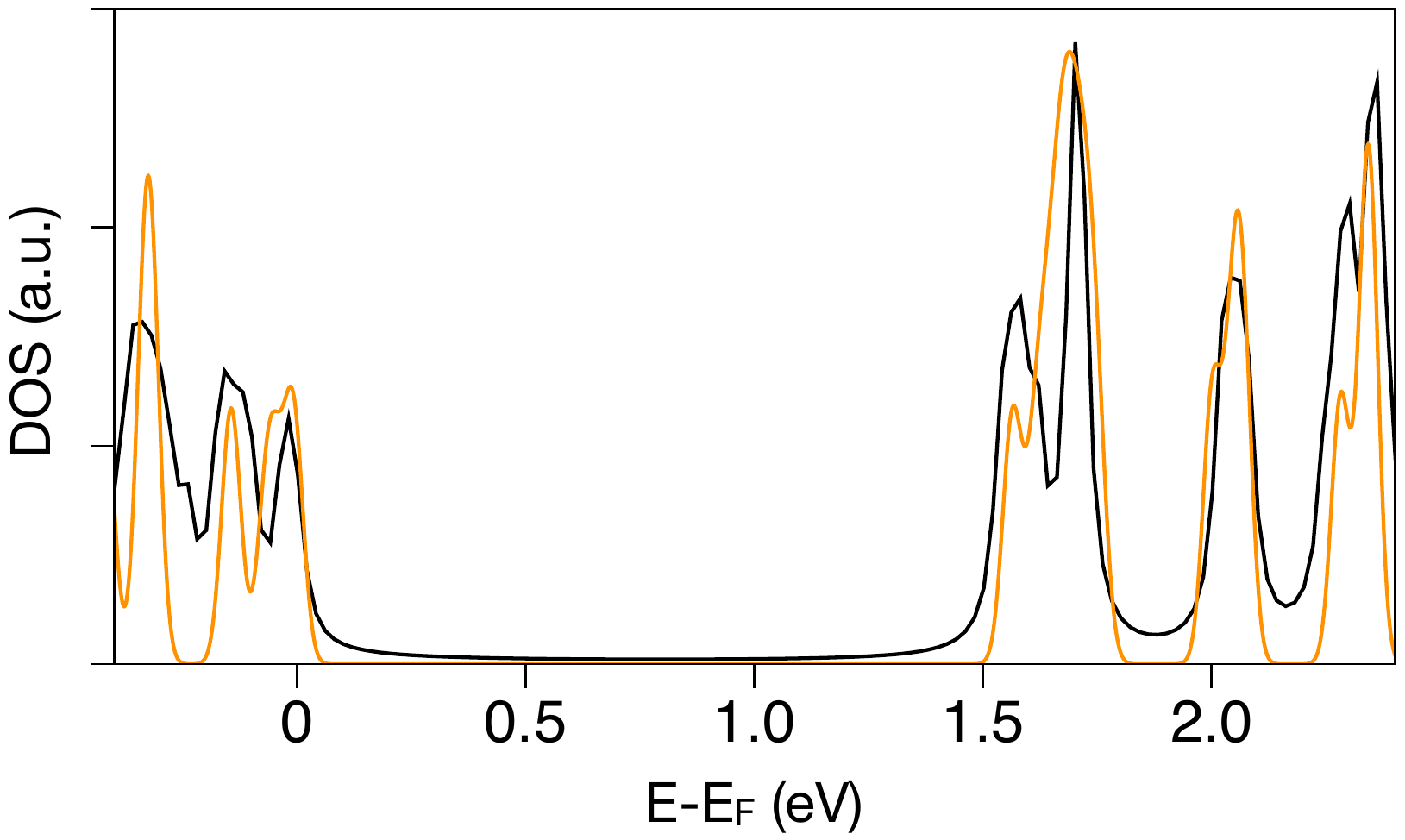}
\caption{(Color online).
Partial density of states (DOS) projected on the bismuth atoms of the structure in Fig. \ref{geometry}. Prediction of DFT is shown in black and that obtained from the tight binding Hamiltonian (omitting the spin-orbit and Rashba terms)  is in orange. }
\label{DOSfit} 
\end{figure}

The tight binding model that we developed previously for the  cowrie shell-like nanostructures, and that is described in detail in Ref.\onlinecite{cowrie}, is applied here to study the properties of nanoparticles consisting of a bismuth monolayer adsorbed on a silicon nanoparticle core, such as that shown in Fig. \ref{geometry}. Both the form of the Hamiltonian and the parameter values in the present work are the same as in Ref.\onlinecite{cowrie} except for the following: 
The site-dependence of the electron's Coulomb potential energy, $H^{\text{C}}_{i}$ in Eq. 2 of Ref. \onlinecite{cowrie}, has been recalculated within DFT for the present system.  The range of the Hamiltonian matrix elements $H^{\text{hop}}$ in Eq. 2 of Ref. \onlinecite{cowrie} that represent electron hopping between Bi atoms has been extended to include interatomic separations greater than 5.5\AA. The values of the parameters $\gamma_n$ in the Hamiltonian term $\tilde{H}^{\text{orb}}$ in Eq. 5 of Ref. \onlinecite{cowrie} have been modified so as to match the Bi partial density of states predicted by the model without spin-orbit and Rashba terms to that predicted by DFT. The Bi density of states obtained from the model in this way and that obtained from DFT can be seen in Fig. \ref{DOSfit}.   

\section{Spin Transport}
\label{ST}

Our calculations of spin filtering by nanoparticles with a bismuth monolayer adsorbed on a silicon core are carried out within the Landauer formalism\cite{Econ81,Fish81} of two-terminal source-drain transport at zero temperature in the linear response regime. We calculate the spin-resolved source to drain electron transmission probabilities by solving the Lippmann-Schwinger equation numerically for source and drain leads attached to individual bismuth atoms of the nanoparticle that is represented by our tight binding Hamiltonian.   The leads are represented by a tight binding model of ideal atomic chains. 
 A detailed description of the model of the leads is given in the last paragraph of Section III of Ref. \onlinecite{cowrie}. We consider spin-unpolarized electrons
entering the nanoparticle through the electron source lead and calculate the spin resolved probabilities $T_{\uparrow}$ and 
$T_{\downarrow}$ of spin up and spin down electrons
exiting through the drain lead at energy $E$. [Note that the transmission probabilities in this paper are Landauer transmission probabilities that are summed over the relevant conducting channels of the source and drain leads and can therefore exceed one.]
We then define the spin polarization
of the electrons entering the drain electrode as 
\begin{equation}\label{Pol}
P=T_{\uparrow}/(T_{\uparrow}+T_{\downarrow}).
\end{equation}
We choose the direction of the axis of quantization for electron spin states in the drain electrode
to be the direction of the expectation value of the spin vector of the electrons carrying the electric current in the drain. This choice ensures that the above definition of the spin polarization $P$ corresponds to the physical spin polarization in the drain lead and is a valid figure of merit for the effectiveness of spin filtering. This methodology of spin transport calculations has been used previously to study spin filtering by our cowrie shell-like bismuthene-silicon bilayer nanostructures and it is described in detail in Ref.\onlinecite{cowrie}.

\begin{figure*}[t]
\centering
\includegraphics[width=1.0\linewidth]{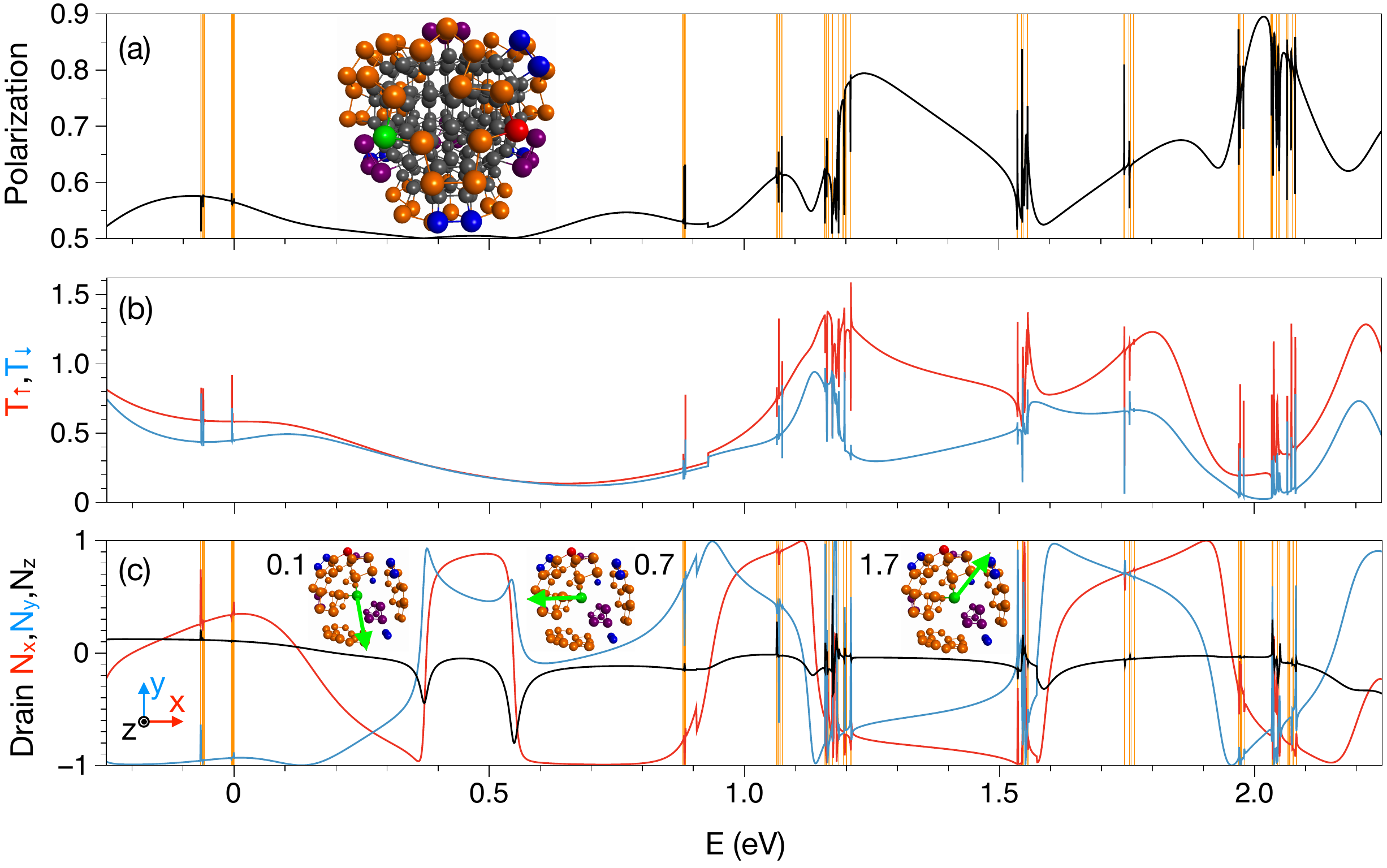}
\caption{(Color online).
Electron spin filtering by the Bi$_{76}$Si$_{147}$ nanostructure shown in Fig.\ref{geometry} when the electron source and drain leads are attached to atoms of the same 14 Bi atom ring. The source and drain Bi atoms are shown in red and green, respectively, in the inset\textcolor{red}{s}. It is assumed that spin-unpolarized
electrons enter the nanostructure from the electron source.  The direction of the axis of quantization for electron spin states in the drain electrode is chosen to be the direction of the expectation value of the spin vector of the electrons carrying the electric current in the drain. (a) The calculated spin polarization $P$ (Eq.\ref{Pol}) vs. electron energy $E$ of electric current exiting the structure to the drain lead is shown in black. The orange vertical lines indicate energy eigenvalues of the tight-binding Hamiltonian when the Bi$_{76}$Si$_{147}$ nanostructure is not connected to the leads. (b)The calculated spin-resolved Landauer transmission probabilities $T_{\uparrow}$ (red) and 
$T_{\downarrow}$ (blue) of spin up and spin down electrons
exiting from the nanostructure into the drain contact at energy $E$. The electron energy is measured from the Fermi level. We note that the strength of the coupling between the Bi$_{76}$Si$_{147}$ nanostructure and the source and drain leads, attested to by the magnitutes of the Landauer transmission probabilities shown here, is sufficent to suppress Coulomb blockade effects. (c)The unit vector $\vec{N}=(N_x, N_y, N_z)$ points in the direction of the spin polarization due to the electric current in the drain lead. $N_x, N_y$ and $N_z$ are plotted in red, blue and black, respectively. The $z$-axis points from the center of the nanoparticle towards the Bi atom (shown in green in the insets) to which the drain lead is attached, and points out of the page in part (c) of the figure. Only the Bi atoms are shown in the insets of part (c). The green arrows show the direction of $\vec{N}$ for selected electron energies that are given in eV next to each schematic. }
\label{filteringonering} 
\end{figure*}

\begin{figure*}[t]
\centering
\includegraphics[width=1.0\linewidth]{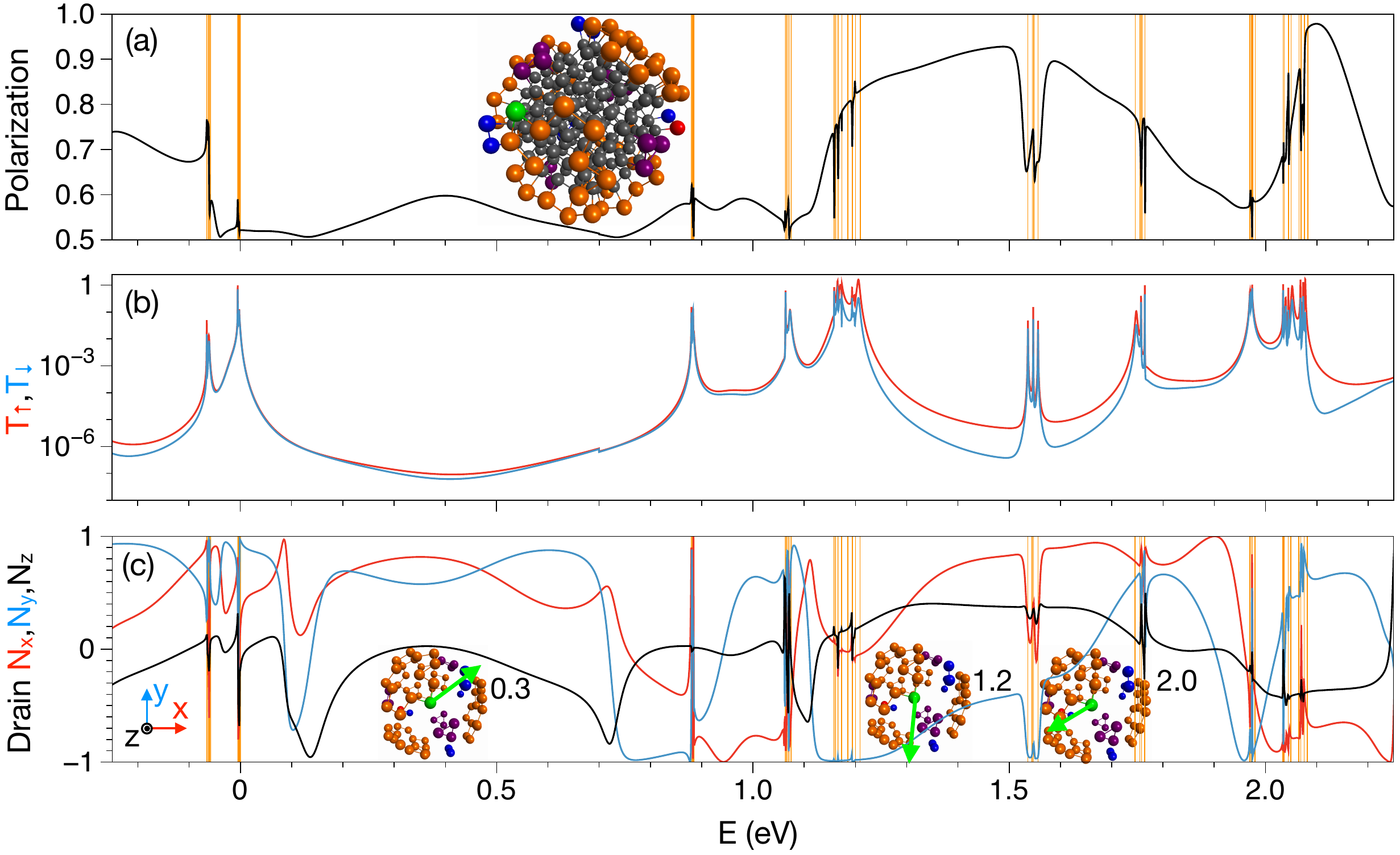}
\caption{(Color online). Electron spin filtering by the Bi$_{76}$Si$_{147}$ nanostructure shown in Fig.\ref{geometry} when the electron source and drain leads are attached to different clusters of Bi atoms. The source lead is attached to an atom (shown in red in the inset) of a Bi atom pair. The drain lead is attached to an atom  (shown in green) of a 14 Bi atom ring.  It is assumed that spin-unpolarized
electrons enter the nanostructure from the electron source. The direction of the axis of quantization for electron spin states in the drain electrode is chosen to be the direction of the expectation value of the spin vector of the electrons carrying the electric current in the drain. (a) The calculated spin polarization $P$ (Eq.\ref{Pol}) vs. electron energy $E$ of electric current exiting the structure to the drain lead is shown in black. The orange vertical lines indicate energy eigenvalues of the tight-binding Hamiltonian when the Bi$_{76}$Si$_{147}$ nanostructure is not connected to the leads. (b)The calculated spin-resolved Landauer transmission probabilities $T_{\uparrow}$ (red) and 
$T_{\downarrow}$ (blue) of spin up and spin down electrons
exiting from the nanostructure into the drain contact at energy $E$. The electron energy is measured from the Fermi level. (c)The unit vector $\vec{N}=(N_x, N_y, N_z)$ points in the direction of the spin polarization due to the electric current in the drain lead. $N_x, N_y$ and $N_z$ are plotted in red, blue and black, respectively. The $z$-axis points from the center of the nanoparticle towards the Bi atom (shown in green in the insets) to which the drain lead is attached, and points out of the page in part (c) of the figure. Only the Bi atoms are shown in the insets of part (c). The green arrows show the direction of $\vec{N}$ for selected electron energies that are given in eV next to each schematic.
}
\label{filteringtwoclust} 
\end{figure*}

\section{RESULTS}
\label{R}

We predict that effective spin filtering by the Bi$_{76}$Si$_{147}$ nanostructure can occur when the source and drain leads are attached to the same Bi atom cluster or to different Bi atom clusters of the nanoparticle, although the transport regimes in the two cases are very different. In the former case conductance of the nanostructure is of the same order of magnitude as that of a metal atomic point contact, i.e., of order $e^2/h$. In the latter case transport requires quantum tunneling between bismuth clusters so that the conductance is typically orders of magnitude lower.

Representative results for an example of the former case, where the source and drain are connected to different atoms of the same 14 member Bi atom ring are shown in Fig.\ref{filteringonering}. The electron source and drain leads are connected to the Bi atoms shown in red and green, respectively in the inset\textcolor{red}{s}. The black curve in Fig.\ref{filteringonering}(a) shows the calculated spin polarization $P$ of electrons emitted into the drain lead, assuming that spin-unpolarized electrons impinge on the nanoparticle from the source lead. The energies of the eigenstates of the tight binding Hamiltonian of the nanoparticle in the absence of leads are shown by the vertical orange lines. In Fig.\ref{filteringonering}(a) strong spin filtering (with $P$ as large 0.895 at $E \sim 2.02$eV) can be seen in the non-resonant transport regime, i.e., when the electron energy does not match that of an eigenstate of the Hamiltonian of isolated nanoparticle. On resonance, where the electron energy does match that of an eigenstate, the spin filtering (i.e., the value of $P$) can be enhanced or reduced relative to nearby non-resonant values. Most resonances are very narrow because they are due to states that occupy multiple Bi clusters that are weakly coupled to each other.  
In Fig.\ref{filteringonering}(b)we show the calculated spin-resolved Landauer transmission probabilities $T_{\uparrow}$ (red) and 
$T_{\downarrow}$ (blue) of spin up and spin down electrons exiting from the nanostructure into the drain contact at energy $E$. These transmission probabilities enter the spin polarization $P$ plotted in Fig.\ref{filteringonering}(a) through Eq. \ref{Pol}. The Landauer conductance of the nanostructure given by $G(E) =\frac {e^2}{h} (T_{\uparrow}+T_{\downarrow})$ has magnitudes in a range typical of a chain of metal atoms, as one might expect for a ring of closely spaced Bi atoms. On resonance, in some cases $T_{\uparrow}$ and $T_{\downarrow}$ are both enhanced or both suppressed or one of them enhanced and the other suppressed, thus giving rise to the resonant enhancement or suppression of the spin polarization $P$ in the drain lead seen in Fig.\ref{filteringonering}(a). 

In  Fig.\ref{filteringonering}(c) we show the unit vector $\vec{N}$ in the direction of the expectation value $\langle\vec{S}\rangle$ of the spin vector of the electrons carrying the the electric current in the drain lead as a function of the electron energy $E$. There $N_x = \langle S_x\rangle/|\langle\vec{S}\rangle|$, $N_y = \langle S_y\rangle/|\langle\vec{S}\rangle|$ and $N_z = \langle S_z\rangle/|\langle\vec{S}\rangle|$ are plotted in red, blue and black, respectively. The $z$-axis points in the direction from the center of the Bi$_{76}$Si$_{147}$ nanostructure towards the (green) bismuth atom to which the drain lead is attached. In the insets of Fig.\ref{filteringonering}(c) (where only the Bi atoms are shown) the $z$-axis points out of the page; the directions of the $x, y$ and $z$ axes in the insets of Fig.\ref{filteringonering}(c) are shown on the left. The green arrows in the insets show explicitly the directions of $\vec{N}$ for a few representative values of the electron energy that are given (in eV units) by the numbers next to the insets. A striking feature of Fig.\ref{filteringonering}(c) is that (except near $E=0.55$eV) $N_z$ is small compared to $(N_x^2+N_y^2)^{1/2}$. This means that for most values of the electron energy the direction of the spin polarization in the drain lead is approximately tangential to the surface of the nanoparticle at the position of the Bi atom to which the drain lead is attached. [That the direction of the spin polarization in the drain lead is approximately tangential to the surface of the nanoparticle appears to be related to the fact that the Bi $6p$ orbital whose symmetry axis is perpendicular to the surface of the nanoparticle is shifted in energy relative to the other Bi $6p$ orbitals due to the interaction between the Bi atom and its silicon nearest neighbor, as is described in Sec. II of Ref. \onlinecite{cowrie}.]  Another striking feature of Fig.\ref{filteringonering}(c) is that direction of the spin polarization in the drain lead (the green arrows in the insets) rotates through {\em large} angles in the $x-y$ plane as the electron energy is varied. Importantly, we conclude that in this system the {\em direction} of the spin polarization in the drain lead can be tuned through {\em large} angles by the varying the electrostatic potential applied to a gate and thus shifting the Fermi energy relative to the electronic energy levels of the  Bi$_{76}$Si$_{147}$ nanoparticle.  

That such purely electrostatic control of the spin polarization direction in the drain electrode (up to and including reversing the direction of the spin polarization by electrostatic means) is at all possible {\em even in principle} is of considerable interest from a fundamental perspective and for potential spintronic device applications.

In Fig.\ref{filteringtwoclust} we present results for an example of the case where the source and drain leads are connected to different Bi atom clusters. As can be seen in the inset, here the electron source lead is connected to a Bi atom (shown in red) of a cluster consisting of 2 Bi atoms whereas the drain lead is connected to a Bi atom (shown in green) of a 14 Bi atom ring. In Fig.\ref{filteringtwoclust}(a), there is very strong non-resonant spin filtering (even stronger than in Fig.\ref{filteringonering}) with the drain spin polarization $P$ as large as 0.979 at $E \sim 2.1$eV. As in Fig.\ref{filteringonering}(a), the spin polarization in Fig.\ref{filteringtwoclust}(a) can be enhanced or weakened on resonance.In Fig\ref{filteringtwoclust}(c) the behavior of $\vec{N}$, the unit vector in the direction of the spin polarization in the drain lead, is qualitatively similar to that in Fig.\ref{filteringonering}(c), although in Fig\ref{filteringtwoclust}(c) $N_z$ exhibits larger deviations from zero than in Fig.\ref{filteringonering}(c).  However, the spin-resolved Landauer transmission probabilities $T_{\uparrow}$ (red) and $T_{\downarrow}$ (blue) in Fig.\ref{filteringtwoclust}(b) are orders of magnitude weaker off resonance than in Fig.\ref{filteringonering}(b). The same is true of the Landauer conductances $G(E) =\frac {e^2}{h} (T_{\uparrow}+T_{\downarrow})$. For instance, at $E \sim 2.1$eV where the spin polarization $P$ is strongest in Fig.\ref{filteringtwoclust}(a), the conductance $G(E) =\frac {e^2}{h}\times1.24\times 10^{-3} $.

It may seem at first sight that such a low conductance would be disadvantageous for spintronic applications. However, this is not necessarily the case: It has been pointed out by Schmidt {\em et al.} \cite{Schmidt} that spin injection from a metal ferromagnet (that has a low resistance) into a semiconductor having a much higher resistance is expected to result in only a very weak spin polarization in the semiconductor. This is because the total resistance of the series circuit of the ferromagnet and semiconductor is dominated by the larger {\em spin-independent} resistance of the semiconductor. A ferromagnet with a much higher  resistance could overcome this difficulty. This line of reasoning suggests that a low conductance (high resistance) configuration of the Bi$_{76}$Si$_{147}$ spin filter may be advantageous for spin injection in some devices. 

Another consideration favoring low conductance configurations of the Bi$_{76}$Si$_{147}$ spin filter is the following: To attach the source and drain leads to {\em different} Bi atom clusters located on {\em opposite} sides of the Bi$_{76}$Si$_{147}$ nanoparticle (as in the low conductance configuration in Fig.\ref{filteringtwoclust}) in either a scanning tunneling microscopy-like setup\cite{STM} or a mechanically controlled break junction\cite{MBJ} would be much more feasible in practice than to attach both the source and drain lead to the same Bi atom cluster. The Bi$_{76}$Si$_{147}$ nanostructure that we consider can be regarded as a molecule of moderate size. Making electrical contact between metal leads and individual atoms at the opposite ends of a single molecule has been accomplished experimentally for a large variety of different molecules.\cite{review} This encourages us to expect experiments such as those that we propose to be possible.

In the situation in Fig. \ref{filteringtwoclust} the electron transmission is low because there is a strong tunnel barrier between the two Bi clusters to which the leads are attached, while each Bi cluster is strongly coupled to its lead; the lead-cluster couplings are as strong in Fig.\ref{filteringtwoclust} as in the system in Fig.\ref{filteringonering}. In this situation pronounced Coulomb blockade effects are not expected because there is only one strong tunnel barrier in the circuit, whereas, as is discussed in Ref. \onlinecite{CB}, significant effects of Coulomb blockade are only observable experimentally when there are at least two strong tunnel barriers in series in a circuit.

In this paper we have presented spin transport results for the case where the source and drain leads are attached to single Bi atoms of the Bi$_{76}$Si$_{147}$ nanoparticle. Making such electrical contact to individual atoms in the laboratory is feasible at the present time with the help of scanning tunneling microscope tips\cite{STM} or nanoscale mechanical break junctions.\cite{MBJ} However, we have also carried out transport calculations with the source and drain leads each connected to multiple Bi atoms of the same Bi cluster or of different Bi clusters. For such arrangements we have also found spin filtering by the Bi$_{76}$Si$_{147}$ nanoparticle.

\section{Conclusions}
\label{Summary} 

In this study we have explored theoretically the electronic structure and spin transport properties of silicon nanoparticles with an adsorbed monolayer of bismuth atoms by means of density functional theory-based calculations and tight binding modeling. In contrast to the previously studied bismuthene-silicon bilayer domes and cowrie shell-like nanostructures in which the bismuth atoms form a single continuous network, the bismuth atoms in the present system are arranged in separate clusters. This clustering of the bismuth atoms strongly affects electronic and spin transport through these nanostructures because the clusters are separated by strong quantum tunnel barriers. We predict strong spin filtering by these nanostructures in the high conductance regime when the source and drain leads are connected to the same bismuth cluster and in the low conductance regime when the source and drain leads are connected to different bismuth clusters. 

    Spin filtering in our non-magnetic Bi$_{76}$Si$_{147}$ nanostructure is made possible by spin-orbit coupling. This is because spin-orbit coupling can result in correlations between the direction in which the spin points and the direction of motion of the electron. A celebrated example of this is the locking between the electron's spin direction and momentum in the edge states of non-magnetic two-dimensional topological insulators where spin up electrons travel in one direction along an edge while spin down electrons travel in the opposite direction.\cite{TIcourse} In our system electrons flow in a specific direction, from the electron source to the drain, and the spin orbit coupling results in a correlation between between the spin direction and the direction of the electric current and hence it results in spin filtering. This and the underlying symmetry of the spin transmission probability matrix are discussed further in the Appendix.

Realization of the present silicon-bismuth nanostructures in the laboratory is expected to be feasible. We expect spin filtering in the low conductance regime to be experimentally accessible and that the low conductance regime may be advantageous for some device applications. We also predict that for such systems the direction of the spin polarization of the electric current that exits from the spin filter into the drain lead can be tuned through large angles and even reversed simply by varying the voltage applied to an electrostatic gate. 
 
\begin{acknowledgments}
This research was supported by NSERC, Westgrid,
and Compute Canada.
\end{acknowledgments}
\appendix
\section{Spin Filtering and the Symmetry of the Spin Transmission Matrix}
\label{TSS'}
\begin{figure*}[t]
\centering
\includegraphics[width=1.0\linewidth]{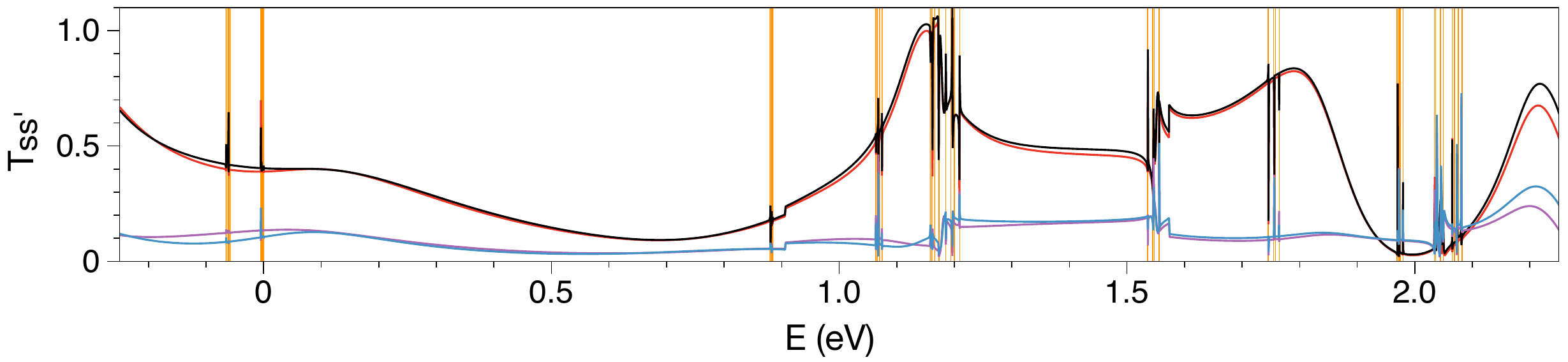}
\caption{(Color online). Spin-dependent electron transmission probabilities  through the Bi$_{76}$Si$_{147}$ nanostructure vs energy E. The source and drain leads are connected to the nanostructure as in Fig.\ref{filteringonering}. Unlike in Fig.\ref{filteringonering}, here the axis of spin quantization points from the center of the nanostructure towards the Bi atom to which the drain lead is attached. $T_{ss'}$ is the Landauer transmission probability of an electron approaching the nanoparticle in the source lead with spin $s'$ and exiting into the drain lead with spin $s$. $T_{\uparrow \uparrow}$, $T_{\downarrow \downarrow}$,$T_{\uparrow \downarrow}$ and $T_{\downarrow \uparrow}$  are shown in red, black, mauve and blue, respectively. The orange vertical lines indicate the energies of the electronic eigenstates of the isolated nanoparticle.
}
\label{spintransmat} 
\end{figure*}

 We find that spin-orbit coupling results in both spin-dependent transmission without spin-flip, and asymmetric spin-flip processes occurring in the transmission of electrons through the Bi$_{76}$Si$_{147}$ nanostructure. The four spin-dependent Landauer transmission probabilities are plotted vs. the electron energy in Fig.\ref{spintransmat} for the same arrangement of source and drain leads as in Fig.\ref{filteringonering}. The axis of spin quantization points from the center of the nanostructure towards the Bi atom to which the drain lead is attached. The differences between the no-spin-flip spin up and spin down transmission probabilities are clearly visible but not large. The two spin-flip transmission probabilities are smaller  and also differ from each other appreciably. Importantly, we find that the $2\times2$ spin transmission probability matrix $T_{ss'}$ is invariant if the source and drain leads are interchanged and simultaneously the spin directions are reversed. This symmetry of the spin transmission probability matrix is related to time reversal symmetry and is obeyed by the correlation between the spin direction and the direction of electron travel through the nanostructure that is responsible for spin filtering.

{

\end{document}
\begin{thebibliography}{333}
{\footnotesize 




\bibitem{Hsu2015}C.-H. Hsu, Z.-Q. Huang, F.-C. Chuang, C.-C. Kuo, Y.-T. Liu, H. Lin, A. Bansil,
The nontrivial electronic structure of Bi/Sb honeycombs on SiC(0001),
New J. Phys. {\bf 17}, 025005 (2015).

\bibitem{Reis2017}F. Reis, G. Li, L. Dudy, M. Bauernfeind, S. Glass, W. Hanke, R. Thomale,
J. Sch\"{a}fer, R. Claessen, Bismuthene on a SiC substrate:
A candidate for a high-temperature
quantum spin Hall material, Science {\bf 357}, 287 (2017). 

\bibitem{Dominguez2018}F. Dominguez, B. Scharf, G. Li, J. Sch\"{a}fer, R. Claessen, W. Hanke, R. Thomale, E. M. Hankiewicz,
Testing Topological Protection of Edge States in Hexagonal Quantum Spin Hall Candidate Materials,
Phys. Rev. B {\bf 98}, 161407(R) (2018).

\bibitem{GLi2018}G. Li, W. Hanke, E. M. Hankiewicz, F. Reis, J. Sch\"{a}fer, R. Claessen, C. Wu, R. Thomale,
Theoretical paradigm for the quantum spin Hall effect at high temperatures,
Phys. Rev. B {\bf 98}, 165146 (2018).

\bibitem{GK2018}G. Kirczenow, Perfect and imperfect conductance quantization and transport resonances of 
two-dimensional topological-insulator quantum dots with normal conducting leads and contacts,
Phys. Rev. B {\bf 98}, 165430 (2018).


\bibitem{Canonico2019}L. M. Canonico, T. G. Rappoport, and R. B. Muniz,
Spin and Charge Transport of Multiorbital Quantum Spin Hall Insulators,
Phys. Rev. Lett. {\bf 122}, 196601 (2019).

\bibitem{Azari2019}M. Azari and G. Kirczenow, Valley polarization reversal and spin ferromagnetism and antiferromagnetism in quantum dots of the topological insulator monolayer bismuthene on SiC
Phys. Rev. B {\bf 100}, 165417 (2019).

\bibitem{Hao2019}X. Hao, F. Luo, S. Zhai, Q. Meng, J. Wu, L. Zhang, T. Li, Y. Jia, M. Zhou,
Strain-engineered electronic and topological properties of bismuthene on SiC(0001) substrate,
Nano Futures {\bf 3}, 045002 (2019).

\bibitem{Stuhler2020}
R. St\"{u}hler, F. Reis, T. Mller, T. Helbig, T. Schwemmer, R. Thomale, J. Sch\"{a}fer, R. Claessen, Tomonaga-Luttinger liquid in the edge channels of a quantum spin Hall insulator, 
Nature Physics {\bf 16}, 47 (2020).

\bibitem{dome}A. Saffarzadeh and G. Kirczenow, Nearly perfect spin filtering in curved two-dimensional topological
 insulators, Phys. Rev. B{\bf 102}, 235420 (2020).
 
 \bibitem{cowrie}A. Saffarzadeh and G. Kirczenow, Resonant and nonresonant spin filtering in bismuthene-silicon 
cowrie shell-like nanostructures, Phys. Rev. B{\bf 104}, 155406 (2021).


\bibitem{Thiessen2019}A. N. Thiessen, M. Ha, R. W. Hooper, H. Yu, A. O. Oliynyk, J. G. C. Veinot, and V. K. Michaelis,
Silicon Nanoparticles: Are They Crystalline from the Core to the Surface?
Chem. Mater. {\bf 31}, 678 (2019).

\bibitem{Huang2021}C.-C. Huang, Y. Tang, M. van der Laan, J. van de Groep, A. F. Koenderink, and K. Dohnalova,
Band-Gap Tunability in Partially Amorphous Silicon Nanoparticles Using Single-Dot Correlative Microscopy,
 ACS Appl. Nano Mater. {\bf 4}, 288 (2021).

\bibitem{Frisch}
M. J. Frisch, G. W. Trucks, H. B. Schlegel, G. E. Scuseria, M. A. Robb, J. R. Cheeseman, G. Scalmani, V. Barone, G. A. Petersson {\em et al.}, the GAUSSIAN 16 Revision: A.03 computer code was used.

\bibitem{MacMolPlt}
B. M. Bode, M. S. Gordon, Macmolplt: a graphical user interface for GAMESS, 
J. Mol. Graphics and Modeling
{\bf 16}, 133 (1998).

\bibitem{SK1954}J. C. Slater and G. F. Koster, Simplified LCAO Method for the Periodic Potential Problem,
Phys. Rev. {\bf 94}, 1498 (1954).

\bibitem{PRBrapid} F. Rostamzadeh Renani and G. Kirczenow, Ligand-Based Transport Resonances of Single-Molecule-Magnet Spin Filters: Suppression of Coulomb Blockade and Determination of Easy-Axis Orientation, Phys. Rev. B {\bf 84}, 180408(R) (2011).

\bibitem{PRB} F. Rostamzadeh Renani and G. Kirczenow, Tight Binding Model of Mn12 Single Molecule Magnets: Electronic and
Magnetic Structure and Transport Properties, Phys. Rev. B {\bf 85}, 245415 (2012).

\bibitem{BR1} Y. A. Bychkov and E. I. Rashba, Properties of a 2D electron gas with lifted spectral 
degeneracy, Pis'ma Zh. Eksp. Teor. Fiz. {\bf 39}, 66 (1984); JETP Lett. {\bf 39}, 78 (1984).
\bibitem{BR2} Y. A. Bychkov and E. I. Rashba, Oscillatory effects and the magnetic susceptibility 
of carriers in inversion layers, J. Phys. C {\bf 17}, 6039 (1984).

\bibitem{Econ81}E. N. Economou and C. M. Soukoulis, Static Conductance and Scaling Theory of Localization in One Dimension,
Phys. Rev. Lett. {\bf 46}, 618 (1981).
\bibitem{Fish81}D. S. Fisher and P. A. Lee, Relation between conductivity and transmission matrix,
Phys. Rev. B {\bf 23}, 6851 (1981).


\bibitem{Schmidt} G. Schmidt, D. Ferrand, L. W. Molenkamp, A. T. Filip and B. J. van Wees, Fundamental obstacle for electrical spin injection from a ferromagnetic metal into a diffusive semiconductor, Phys. Rev. B {\bf 62}, R4790 (2000).

\bibitem{STM}M. F. Crommie, C. P. Lutz, D. M. Eigler, 
Confinement of Electrons to Quantum Corrals on a Metal Surface,
Science, {\bf 262}, 218 (1993). 
\bibitem{MBJ}M. A. Reed, C. Zhou, C. J. Muller, T. P. Burgin, J. M. Tour,
Conductance of a Molecular Junction,
Science {\bf 278}, 252 (1997).

\bibitem{review}{G. Kirczenow, in The Oxford Handbook of Nanoscience and Technology, Volume I: Basic Aspects, edited by A. V. Narlikar and Y. Y. Fu (Oxford University Press, Oxford, 2010), Chap. 4.}

\bibitem{CB}{H. Grabert,
Single Charge Tunneling: A Brief Introduction,
Z. Phys. B - Condensed Matter {\bf 85}, 319 (1991).}


\bibitem{TIcourse}J. K. Asb\'{o}th, L. Oroszl\'{a}ny, A. P\'{a}lyi,
Sec. 8.4, {\em A Short Course on Topological Insulators}, 
Vol. 919, {\em Lecture Notes in Physics}, 
Springer, 2016. DOI 10.1007/978-3-319-25607-8



}
\end{thebibliography}
